\documentclass[epj]{webofc}
\usepackage[varg]{txfonts}   
\usepackage{subfigure}
\usepackage{graphicx}
\newcommand{\be}{\begin{equation}}
\newcommand{\ee}{\end{equation}}
\newcommand{\bea}{\begin{eqnarray}}
\newcommand{\eea}{\end{eqnarray}}
\newcommand{\nn}{\nonumber}

\begin{document}
\title{Extended Higgs sectors in the context of supersymmetry at the LHC }
%
%
{IITH-PH-0002/16}
\author{\firstname{Priyotosh} \lastname{Bandyopadhyay}\inst{1}\fnsep\thanks{\email{priyotosh.bandyopadhyay@le.infn.it, bpriyo@iith.ac.in}}}

\institute{Dipartimento di Matematica e Fisica "Ennio De Giorgi", \\ Universit\`a del Salento and INFN-Lecce, Via Arnesano, 73100 Lecce, Italy\\ and \\
Indian Institute of Technology Hyderabad, Kandi,  Sangareddy-502287, Telengana, India}

\abstract{We studied the extension of Standard Model (SM) Higgs sector with $SU(2)$ triplet and SM gauge singlet in the context of supersymmetry. For a $Z_3$ symmetric superpotential where the singlet field generates the bi-linear Higgs mixing term between two Higgs doublets dynamically. Models like this often have a very light pseudoscalar as pseudo-Nambu Goldstone model of an extra $U(1)$ symmetry. We investigate the possibility of such light hidden Higgs boson with the data of the discovered Higgs boson around 125 GeV. Along with the neutral sector, the charged Higgs bosons in the triplet extended case are really interesting as the triplet charged Higgs bosons give rise to new decay mode to $Z W^\pm$  along with $ a_1 W^\pm$ due to the existence of light pseudoscalar. These new modes can be looked for at the LHC in the search for the Higgs bosons in other representations of $SU(2)$. }
\maketitle
\section{Introduction}
The discovery of the Higgs boson around 125 GeV has certainly proved the existence of at least one scalar taking part in the process of electroweak symmetry breaking (EWSB) \cite{CMS}. The presence of such a scalar has been experimentally verified in its decay modes into $WW^*$,
$ZZ^*$ and $\gamma\gamma$ \cite{CMS} at the Large Hadron Collider (LHC).  In the midst of searching for SM decay modes, it is also important to investigate other non standard searches which are often motivated by various theoretical constructs. 

In this article we are going to illustrate the theoretical possibilities which remain, at this time, wide open from a phenomenological perspective and which define the possible scenarios for an extended Higgs sector, in the context of supersymmetry. These constructions also address other theoretical issues, with exciting phenomenological implications. The extension of the Higgs sector with a singlet superfield can solve the $\mu$ problem in the minimal supersymmetric SM (MSSM) dynamically.  Similarly one can introduce an $SU(2)$ triplet superfield, but in this case such dynamical solution is not possible, due to the constraints imposed by the $\rho$ parameter on the vacuum expectation value (v.e.v.) of the triplet \cite{rho}. Extending the Higgs sector by a singlet/triplet also gives the possibility of a spontaneous violation of the CP symmetry.

We consider a supersymmetric extension of the Higgs sector by a $SU(2)\times U(1)_Y$ singlet and a $Y=0$ hypercharge $SU(2)$ triplet. Additional $Z_3$ symmetry of the Lagrangian predicts the existence of a light pseudoscalar in some limit of the theory. Such possibility leads to additional decay modes of the Higgs boson around 125 GeV into a pair of light pseudoscalars and/or into a light pseudoscalar accompanied by a $Z$ boson, if kinematically allowed.

The charged Higgs sector is extended due to the triplet but they are fermiophobic leading to distinguishing features. A light charged Higgs boson may now decay into a pseudoscalar and one $W^\pm$ boson. This particular decay is allowed for both doublet and triplet-like charged Higgs bosons. However, the $Y=0$ triplet brings in an additional coupling of the charged Higgs boson to the $Z$ and $W^\pm$ at tree-level, which breaks the custodial symmetry. This can generate a totally new decay mode for the charged Higgs boson into $Z\, W^\pm$, which is not present in the 2-Higgs doublet model (2HDM) or in the MSSM. Here we are going to discuss such possibilities and elaborate on how to explore them at current and future experiments

In section~\ref{Model} we are going to briefly illustrate the model.  In section~\ref{lipseudo} we address the possibility of having a light pseudoscalar and its phenomenology. The charged Higgs boson phenomenology has been discussed in section~\ref{charged} with conclusions in section~\ref{concl}. 
\section{Model}\label{Model}
As explained in \cite{TNSSMo} the model  contains a SU(2) triplet $\hat{T}$ of zero hypercharge ($Y=0$)  together with a SM gauge singlet ${\hat S}$, added to the superfield content of the MSSM,
\begin{equation}
 W_{TNMSSM}=W_{MSSM} + W_{TS}.
 \end{equation}
The structure of its triplet and singlet extended superpotential can be written as 

 \begin{equation}
W_{TS}=\lambda_T  \hat H_d \cdot \hat T  \hat H_u\, + \, \lambda_S \hat S\,  \hat H_d \cdot  \hat H_u\,+ \frac{\kappa}{3}\hat S^3\,+\,\lambda_{TS} \hat S \, \textrm{Tr}[\hat T^2] .
\label{spt}
 \end{equation}
 The triplet and doublet superfields are given by 
 \begin{equation}\label{spf}
 \hat T = \begin{pmatrix}
       \sqrt{\frac{1}{2}}\hat T^0 & \hat T_2^+ \cr
      \hat T_1^- & -\sqrt{\frac{1}{2}}\hat T^0
       \end{pmatrix},\qquad \hat{H}_u= \begin{pmatrix}
      \hat H_u^+  \cr
       \hat H^0_u
       \end{pmatrix},\qquad \hat{H}_d= \begin{pmatrix}
      \hat H_d^0  \cr
       \hat H^-_d
       \end{pmatrix}.
 \end{equation}
 Here $\hat T^0$ denotes a complex neutral superfield, while  $\hat T_1^-$ and $\hat T_2^+$ are the charged Higgs superfields.  The MSSM Higgs doublets are the only superfields which couple to the fermion multiplet via Yukawa coupling. All the coefficients involved in the Higgs sector are chosen to be real in order to preserve CP invariance. The breaking of the $SU(2)_L\times U(1)_Y$ electroweak symmetry is then obtained by giving real vevs to the neutral components of the Higgs field
 \be
 <H^0_u>=\frac{v_u}{\sqrt{2}}, \, \quad \, <H^0_d>=\frac{v_d}{\sqrt{2}}, \quad <S>=\frac{v_S}{\sqrt{2}}, \, \quad\, <T^0>=\frac{v_T}{\sqrt{2}},
 \ee
 which give mass to the $W^\pm$ and $Z$ bosons
 \be
 m^2_W=\frac{1}{4}g^2_L(v^2 + 4v^2_T), \, \quad\ m^2_Z=\frac{1}{4}(g^2_L \, +\, g^2_Y)v^2, \, \quad v^2=(v^2_u\, +\, v^2_d), 
\quad\tan\beta=\frac{v_u}{v_d} \ee
 and also induce, as mentioned above, a $\mu$-term of the form $ \mu_D=\frac{\lambda_S}{\sqrt 2} v_S+ \frac{\lambda_T}{2} v_T$. The triplet vev $v_T$ is strongly  constrained by the global fit to the measured value of the $\rho$ parameter \cite{rho}  which restricts its value to $v_T \leq 5$ GeV  and in the numerical analysis we have chosen $v_T =3$ GeV.
 
 In the TNMSSM,  we are left with four CP-even, three CP-odd  and three charged Higgs bosons as shown below
 \begin{align}\label{hspc}
\rm{CP-even} &\quad \quad  \rm{CP-odd} \quad\quad   \rm{charged}\nn \\
 h_1, h_2, h_3, h_4 &\quad \quad a_1, a_2, a_3\quad \quad h^\pm_1, h^\pm_2, h^\pm_3. \nn
 \end{align}
The neutral Higgs bosons are linear combinations of doublets, triplet and singlets, whereas the charged Higgses are  combinations of doublets and of a triplet only. At tree-level the maximum value of the lightest neutral Higgs has additional contributions from the triplet and the singlet sectors respectively. The upper bound on the lightest CP-even Higgs can be extracted from the relation
\be\label{hbnd}
m^2_{h_1}\leq m^2_Z(\cos^2{2\beta} \, +\, \frac{\lambda^2_T}{g^2_L\,+\,g^2_Y }\sin^2{2\beta}\, +\, \frac{2\lambda^2_S}{g^2_L\,+\,g^2_Y }\sin^2{2\beta}),
\ee
which is affected on its right-hand-side by two additional contributions from the triplet and the singlet. These can raise the allowed tree-level Higgs mass and additional contributions coming from the radiative corrections of the  triplet and the singlet reduce the required the supersymmetric (SUSY) mass scale and thus the fine-tuning problem. We have calcualted the one-loop Higgs mass for the neutral Higgs bosons  in \cite{TNSSMo} following  the Coleman-Weinberg effective potential \cite{Coleman:1973jx}. The existence of the triplet and singlet fields at quantum level are non-negligible which reduce the SUSY fine-tuning for all $\tan{\beta}$ values \cite{TNSSMo, pbas1, pbas2}.

\section{Light pseudoscalar and phenomenology}\label{lipseudo}
 One of the most interesting feature can be seen  in the limit where the trilinear soft parameters $A_i$ in go to zero, the  discrete $Z_3$ symmetry of the Lagrangian is promoted to a continuos $U(1)$ symmetry given by Eq.~(\ref{csmy}) 
\begin{align}\label{csmy}
(\hat{H}_u,\hat{H}_d, \hat{T},\hat{ S}) \to e^{i\phi}(\hat{H}_u,\hat{H}_d, \hat{T},\hat{ S}) .
\end{align}
An explicit breaking of a continuous global symmetry is expected to be accompanied by pseudo-Nambu Goldstone bosons (pNGB) as in the case of chiral symmetry breaking in QCD where the pions take the role of the corresponding pNGB. In this case, in the TNMSSM we should expect a light pseudoscalar in the spectrum, whose mass is of the same order of the $A_i$ parameters. A similar behaviour is allowed in the  NMSSM and such light pseudoscalar is known as the $R$-axion \cite{nmssm, Agashe:2011ia}.   However, such light pseudoscalar receives bounds if it decays into fermions.  Previous analysis at LEP have searched for such light scalar bosons in
$e^+ e^-\to Z h$ and $e^+ e^-\to A h$, where $h, A$ are CP-even and odd neutral Higgs bosons respectively \cite{LEPb}. Similar bounds also come from the bottomonium decay \cite{bottomium} for such light pseudoscalar in the mass range of 5.5 -14 GeV. Recently published data from CMS also provide strong bounds for such a light pseudoscalar when it couples to fermions \cite{cmsab}. Being mostly singlet-like, it is easy to evade such bounds but difficult to produce it directly at a hadron collider. 

\begin{figure}[h]
\begin{center}
\mbox{\subfigure[]{\hskip -15 pt
\includegraphics[width=0.45\linewidth]{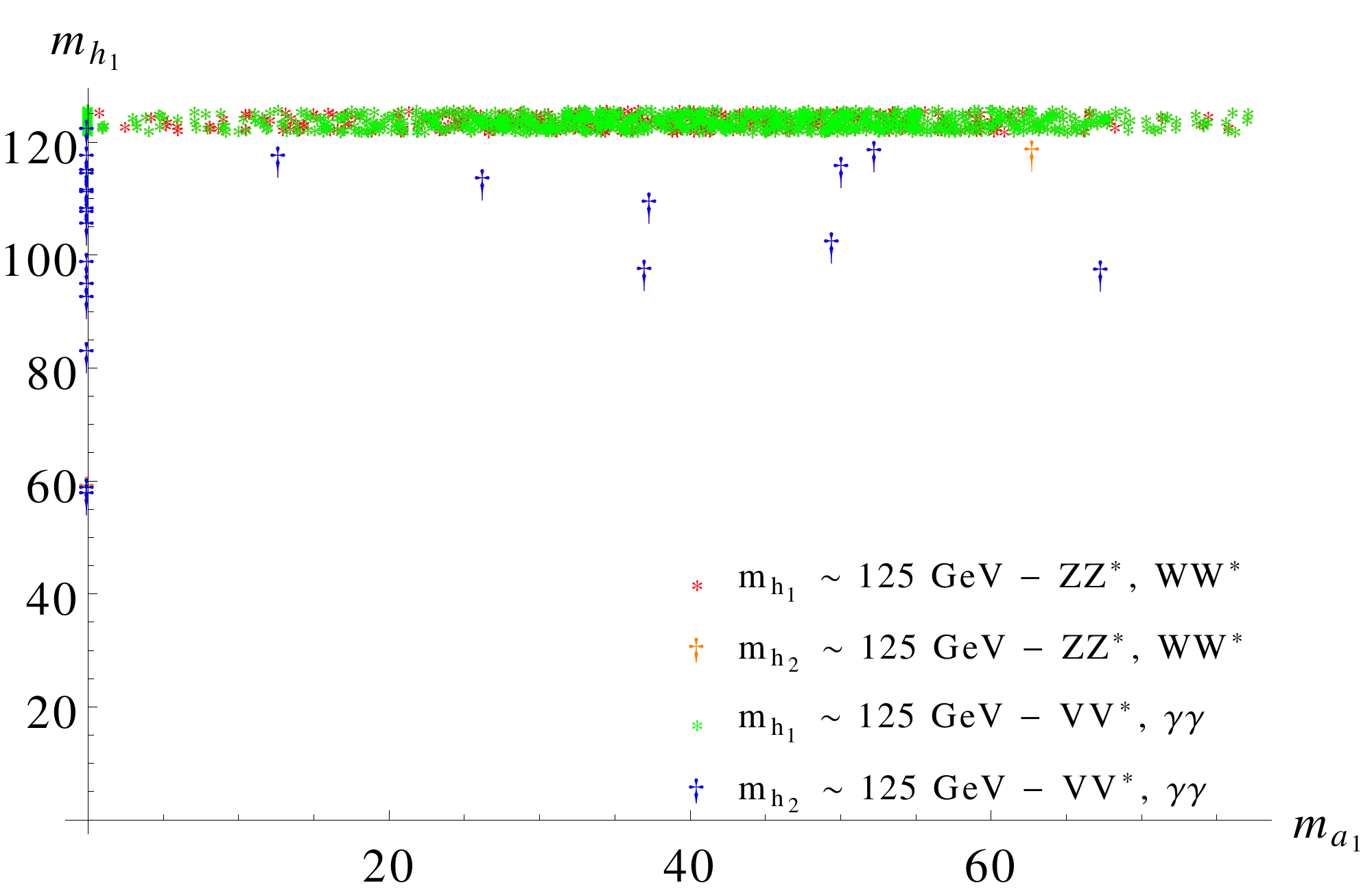}}
\subfigure[]{\includegraphics[width=0.45\linewidth]{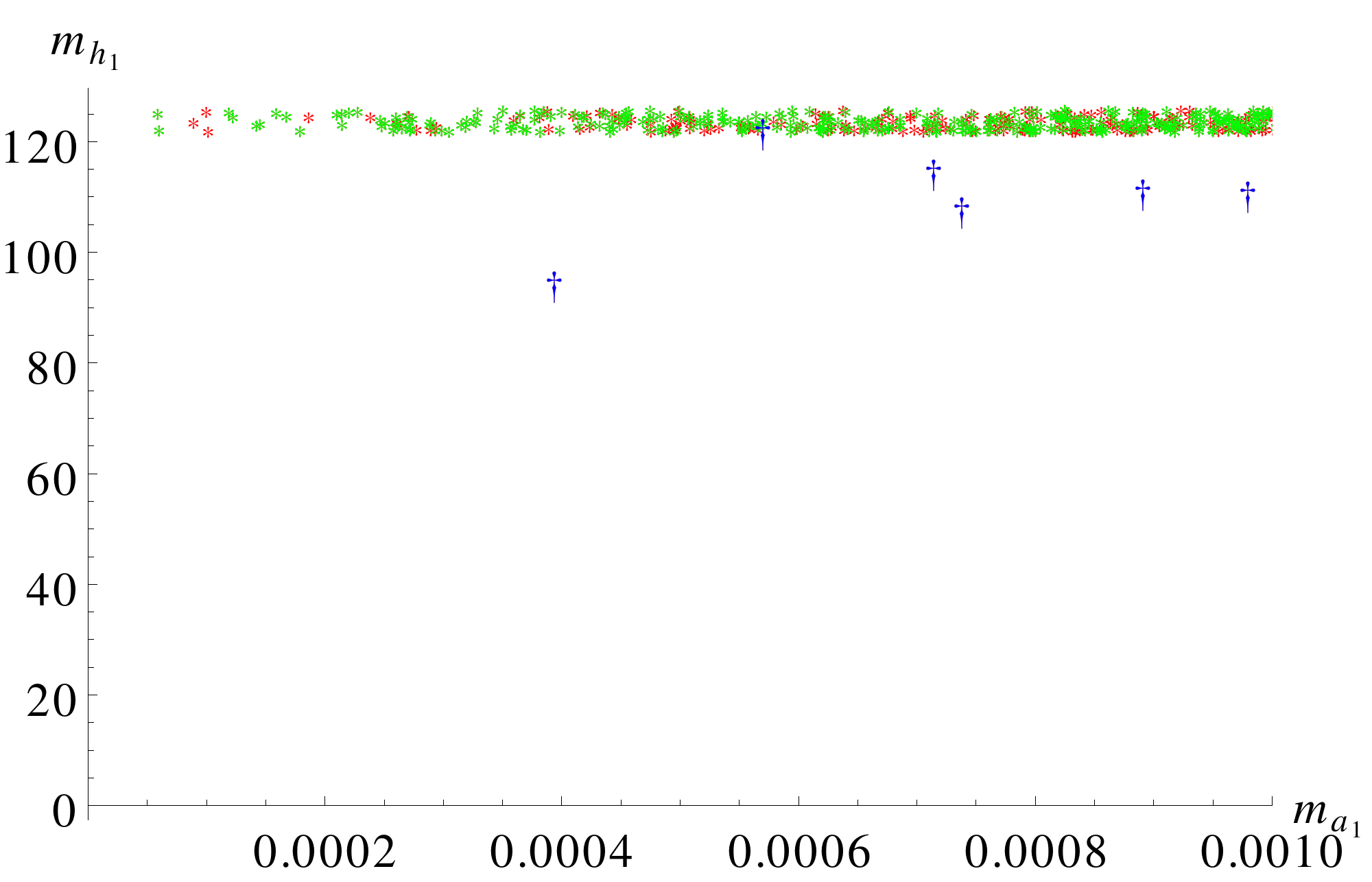}}}
\caption{The lightest CP-even Higgs boson mass $m_{h_1}$ vs the lightest pseudoscalar mass $m_{a_1}$ at one-loop consistent with the Higgs data from CMS, ATLAS and LEP.  Very light pseudoscalar masses $m_{a_1}\leq 1$ MeV are shown in panel (b)\cite{TNSSMo}.}\label{higgsdata}
\end{center}
\end{figure}

\begin{figure}[hbt]
\begin{center}
\includegraphics[width=0.4\linewidth]{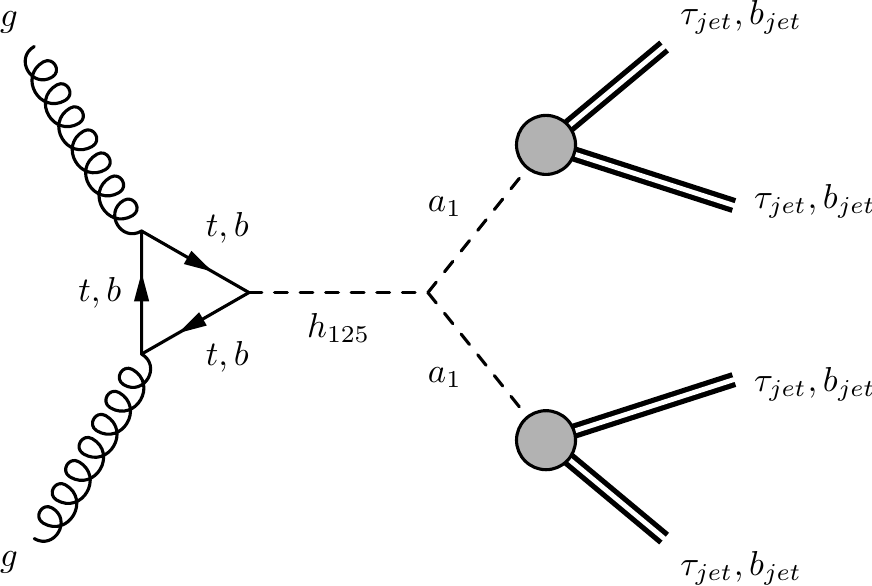}
\caption{Pseudoscalar (triplet/singlet) pair production from Higgs boson, generated via 
gluon-gluon fusion, and their decays via their mixing with the doublets \cite{TNMSSM2}.}\label{glutri}
\end{center}
\end{figure}

In Figure~\ref{higgsdata} we show a Higgs $\sim 125$ GeV which at the same time satisfy the $ZZ^*$, $WW^*$ bounds at $1\sigma$ level and the $\gamma\gamma$ bound at $2\sigma$ level from both CMS and ATLAS along with a light pseudoscalar with or without a light scalar. Such singlet type light pseudoscalar couples to the doublet-like Higgs bosons via $\lambda_S$ (see Eq.~\ref{spt}). This makes it easy to produce by an intermediate Higgs. If such a light pseudoscalar ($m_{a_1}\le 125/2$) exists, then it can be produced in $gg \to h_{125} \to a_1 a_1$ as shown in Figure~\ref{glutri}.  Such pair production can give rise to final states rich in $\tau$, $b$ and even in muons. In \cite{TNMSSM2} we have investigated such final states by considering all the dominant SM backgrounds at the LHC with a center of mass energy of 13 and 14 TeV. A detailed signal to background analysis of the final states $2\tau + 2b$, $2b +2\mu$ and  $\geq 3\tau$ reveals that some of the benchmark points with the light pseudoscalar ($m_{a_1}\sim 10-20$ GeV) can be probed with early data of 25 fb$^{-1}$ at the LHC.

\section{Charged Higgs boson Phenomenology}\label{charged}
As mentioned earlier that unlike in NMSSM and MSSM, TNMSSM has three charged Higgs bosons out of which two of them are of triplet in nature which do not couple to fermions and that makes it hard to produce them directly. In the mass basis they are generally in mixed states. The presence of a light pseudoscalar  affects the charged Higgs decay phenomenology and such light state can be probed via $h^\pm_1 \to a_1 W^\pm$. Similarly in the case of  NMSSM with a $Z_3$ symmetry \cite{nmssm} the non-standard decay mode of the light charged Higgs boson $h^\pm \to a \,W^\pm$ can evade the recent mass bounds from LHC \cite{chLHC} and can be probed with the $\tau$ and $b$ rich final states coming from the light pseudoscalar \cite{chaW}. The charged Higgs boson in this case (NMSSM) is still doublet-like so its production from the decay of a top or via $b g\to t h^\pm$ is still possible. The possibility of a light scalar 
is also present in the MSSM with CP-violating interactions, where the light Higgs is mostly CP-odd and can evade the existing LEP bounds. Such Higgs boson with a mass $\lesssim 30$ GeV can open the decay mode $h^\pm \to h_1 W^\pm$,  which can be explored at the LHC\cite{cpv}.

\begin{figure}[t]
\begin{center}
\mbox{\subfigure[]{\includegraphics[width=0.2\linewidth]{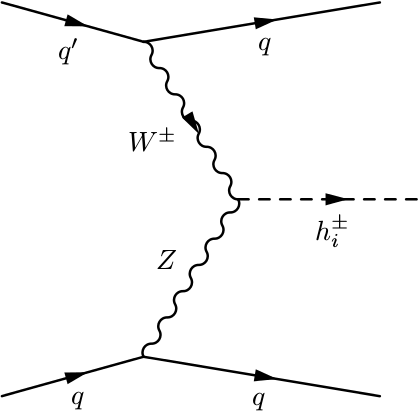}}
\hskip 25 pts
\subfigure[]{\includegraphics[width=0.45\linewidth]{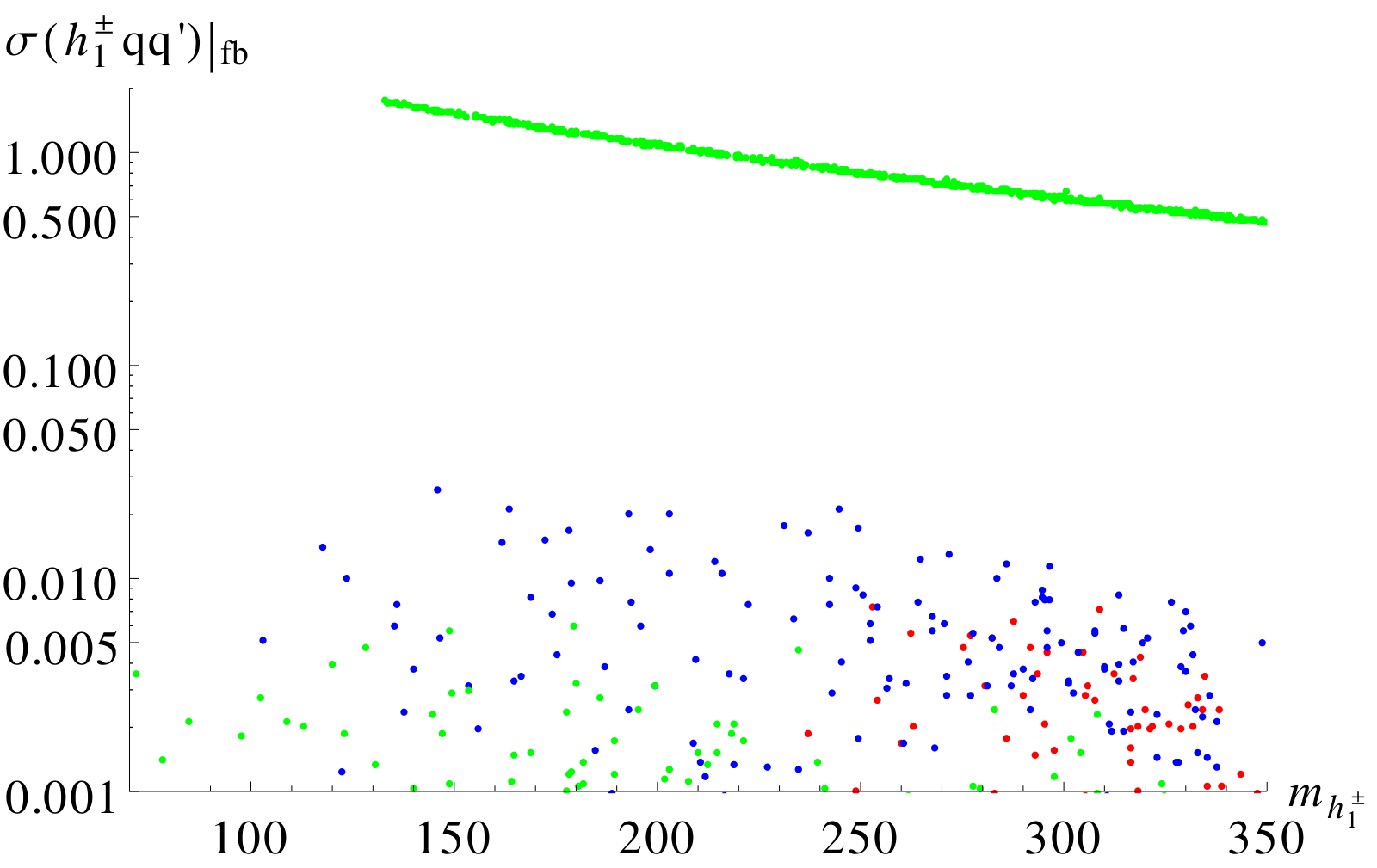}}}
\caption{The Feynman diagram for the charged Higgs production via vector boson fusion at the LHC (a)
and the production cross-section of a light charged Higgs boson via  vector boson fusion versus the light charged Higgs boson mass $m_{h^\pm_1}$ (b) \cite{tripch}.}\label{prodvvfch}
\end{center}
\end{figure}

However things can change a lot in a triplet extended supersymmetric model (TESSM), where we have a light charged Higgs triplet. It does not couple to fermions, which makes its production via conventional methods a lot harder. However, its production channels in association with $W^\pm$ bosons or with neutral Higgs bosons are still considerable. The interesting feature of having a charged Higgs triplet is the non-zero $h^\pm_i - Z- W^\mp$ coupling, which not only gives rise to a new decay mode of $ZW^\pm$ but also makes the charged Higgs production possible via vector boson fusion \cite{tripch}  as can be seen from Fig~\ref{prodvvfch}. 
This is not allowed for the doublet-like charged Higgs boson of the MSSM or of the 2HDM. In the TNMSSM we have both the presence of a very light pseudoscalar - as in the NMSSM -  and of a charged Higgs triplet, as in the TESSM. If we can pair produce such charged Higgs where one of them decays into $a_1 W^\pm$ and the other into $ZW^\pm$, then we can search for such decay modes. This provides a smoking gun
signature for the extended Higgs structure of the TNMSSM.  We are currently investigating in detail ways to probe both aspects at the LHC \cite{Chsim}.

\section{Conclusions}\label{concl}
The discovery of a Higgs boson around 125 GeV in mass cannot explain all the shortcomings of SM
and the quest for extra Higgs bosons are still there. Supersymmetric theories additional Higgs bosons which may belong to same or higher representations of the $SU(2)$.  Phenomenologically the existence of a very light scalar below 100 GeV is still there and keeps the theoretical possibility of various representations. Given that all the SM decay modes of the Higgs boson around $125$ GeV are yet to be discovered, the question of the existence of non-standard decays in this sector still remains elusive. The other Higgs bosons could have higher mass values  ($> 125$ GeV) but the possibilities of lower mass values ($< 100$ GeV) also remain quite open. Certainly in such cases, if the $h_{125}\to h_1h_1/a_1a_1$ decay modes are 
kinematically allowed, then they can be probed at the LHC. 
Models with $Z_3$ symmetry, such as the NMSSM or the TNMSSM, where such scalars are naturally light, can be tested once additional data will be made available from the LHC. If an extended Higgs sector exists, then finding a charged Higgs is a direct proof of it. So far, the search for a charged Higgs boson is performed in its decay to $\tau\nu$, but the existence of a light scalar/pseudoscalar gives us the new decay mode $h^\pm \to a_1/h_1 W^\pm$, which should be looked for at the LHC. Triplet charged and neutral Higgs bosons do not couple to fermions and a charged Higgs triplets decay to $ZW^\pm$ which can be explored at the LHC. To complete our understanding of electroweak phase transition we need to explore all the theoretical possibilities experimentally.

\section{Acknowledgement} 
PB wants to thank University of Salento  for the invitation and support to attend the workshop.

\end{document}